# Structural changes at the semiconductor-insulator phase transition in the single layered $La_{0.5}Sr_{1.5}MnO_4$ perovskite.

Javier Herrero-Martín[1,2], Javier Blasco[3], Joaquín García[3], Gloria Subías[3], and Claudio Mazzoli[1]

[1] *European Synchrotron Radiation Facility, F-38043 Grenoble Cedex 9, France*

[2] *Institut de Ciència de Materials de Barcelona, CSIC, Campus Universitari de Bellaterra, E-08193 Bellaterra, Spain*

[3] *Instituto de Ciencia de Materiales de Aragón, Departamento de Física de la Materia Condensada, CSIC–Universidad de Zaragoza, E-50009 Zaragoza, Spain*

**PACS numbers:** 75.47.Lx, 78.70.Ck, 75.50.Ee, 61.10._i

**Corresponding author:**

Joaquín García Ruiz

Email: jgr@unizar.es

# ABSTRACT

The semiconductor-insulator phase transition of the single-layer manganite $La_{0.5}Sr_{1.5}MnO_4$ has been studied by means of high resolution synchrotron x-ray powder diffraction and resonant x-ray scattering at the Mn K edge. We conclude that a concomitant structural transition from tetragonal *I4/mmm* to orthorhombic *Cmcm* phases drives this electronic transition. A detailed symmetry-mode analysis reveals that condensation of three soft modes —$\Delta_2$(B2u), $X_1^+$(B2u) and $X_1^+$(A)— acting on the oxygen atoms accounts for the structural transformation. The $\Delta_2$ mode leads to a pseudo Jahn-Teller distortion (in the orthorhombic *bc*-plane only) on one Mn site (Mn1) whereas the two $X_1^+$ modes produce an overall contraction of the other Mn site (Mn2) and expansion of the Mn1 one. The $X_1^+$ modes are responsible for the tetragonal superlattice (1/2,1/2,0)-type reflections in agreement with a checkerboard ordering of two different Mn sites. A strong enhancement of the scattered intensity has been observed for these superlattice reflections close to the Mn K edge, which could be ascribed to some degree of charge disproportion between the two Mn sites of about 0.15 electrons. We also found that the local geometrical anisotropy of the Mn1 atoms and its ordering originated by the condensed $\Delta_2$ mode alone perfectly explains the resonant scattering of forbidden (1/4,1/4,0)-type reflections without invoking any orbital ordering.

# I. INTRODUCTION

Mixed-valence manganese perovskites have been matter of intensive research due to their remarkable interrelated physical properties.[1,2] Depending on composition, they show a wide variety of magnetic and electrical phenomena, including ferromagnetic-metal, antiferromagnetic-insulating and charge and orbital orderings. Colossal magnetoresistance behavior has been observed in the $RE_{1-x}(Sr/Ca)_xMnO_3$ perovskite and the double-layered $La_{2-2x}Sr_{1+2x}Mn_2O_7$, but not in the single-layered system $La_{1-x}Sr_{1+x}MnO_4$[3,4]. This indicates that the presence of magnetoresistance depends not only on the competition among the charge, orbital and spin degrees of freedom of Mn but also on the lattice dimensionality.

The physics of mixed valence oxides of 3$d$ transition metals is determined by the nature of charge carriers. Traditionally, mixed valence manganites have been described as ionic compounds in such a way that a formal valence of the manganese $v_{Mn}=3+\delta$ was considered to be composed by a bimodal distribution of (1-$\delta$) $Mn^{3+}$ and $\delta Mn^{4+}$ ions. In this way, the semiconductor-insulator transition observed on decreasing temperature in half doped manganites ($\delta=0.5$) has been interpreted as a transition from a system with delocalized electrons (metallic phase) to a periodic ordering of $Mn^{3+}$ and $Mn^{4+}$ ions in the lattice, *i.e.* a charge ordering (CO). Since the electronic configuration for $Mn^{3+}$ is $t_{2g}^3 e_g^1$, active to develop a Jahn-Teller (J-T) distortion, a simultaneous ordering in the orientation of the $e_g$ orbitals coupled to the CO was proposed, giving rise to the so-called orbital ordering (OO). The insulating phase in half-doped manganites has long been described as found in the Goodenough model of a checkerboard pattern of $Mn^{3+}$ and $Mn^{4+}$ ions with the occupied $Mn^{3+}$ $e_g$ orbitals *zigzag* ordered in the CE-type antiferromagnetic pattern[5]. This CO-OO picture seems to be confirmed by x-ray and neutron diffraction experiments in the three dimensional $La_{0.5}Ca_{0.5}MnO_3$ perovskite, where a checkerboard ordering of crystallographic inequivalent Mn sites was observed[6]. A pioneer resonant x-ray scattering (RXS) experiment by Murakami *et al.* on the layered $La_{0.5}Sr_{1.5}MnO_4$ perovskite apparently agrees with the above picture.[7] Nevertheless, after this experiment, numerous RXS works at the Mn K edge in manganites have reported on a nearly indiscernible charge disproportionation between the two different crystallographic sites for the Mn atom.[8-12] Moreover, Bragg-forbidden reflections ascribed to OO are mainly originated by local structural distortions, hardly conciliable to the J-T effect in a $Mn^{3+}$ ion.[8-16] In recent years, soft RXS at the Mn $L_{2,3}$ edges has been proposed as the unique technique for the direct observation of orbital and spin correlations in manganites as it directly probes the $3d$ states. Consequently a large number of soft RXS studies are mostly interpreted in terms of the $e_g$ OO yet.[17-25] In light of the latter considerations, we re-examine here the case of the layered $La_{0.5}Sr_{1.5}MnO_4$ manganite, which is considered as one of the prototypical compounds displaying a CO-OO state, combining RXS at the Mn K edge and high-resolution synchrotron powder x-ray diffraction.

$La_{0.5}Sr_{1.5}MnO_4$ belongs to the $n=1$ members of the Ruddlesden-Popper family of manganites of general formula $La_{n(1-x)}Sr_{nx+1}Mn_nO_{3n+1}$, in which there is a periodic arrangement of consecutive layers of $(La,Sr)MnO_3$ perovskite and $(La,Sr)O$ rock-salt-type along the $c$-axis. This compound does not show colossal magnetoresistance and its layered structure results in strongly anisotropic transport properties.[26,27] The crystal structure at room temperature is tetragonal with space group *I4/mmm* and lattice constants $a$=3.86 Å and $c$=12.42 Å. At about $T_{CO}$=230 K, the sample undergoes a structural phase transition also considered as a CO phase transition due to the sharp discontinuity in the electrical resistivity. The magnetic structure is CE-type with the Néel temperature $T_N$~110 K.[28]

The semiconductor-insulator transition has been studied by several techniques. The low temperature phase was first observed by electron microscopy, which revealed the presence of superstructure peaks below 230 K following the $(1/4,1/4,0)_t$ wave vector modulation of the room temperature structure (hereafter, the subscript t refers to the tetragonal cell).[26,27] Sternlieb *et al.*[28] confirmed the existence of a distinct low-temperature phase by a neutron scattering experiment but they concluded that this phase has a higher structural symmetry with $(1/2,1/2,0)_t$ wavevector periodicity. The more complete study performed by Larochelle *et al.*,[29,30] showed weak $(1/2,1/2,l)_t$ and $(1/4,1/4,l)_t$ reflections, leading to a unit cell of the type $\sqrt{2}a_t \times 2\sqrt{2}a_t \times c_t$. In this paper, the low temperature phase is described in the space group *Bbmm* (standard setting *Cmcm*), but the complete refinement of the crystallographic structure was not performed and the authors suggest a non-centrosymmetric group (*B2mm*). On the other hand, Senff *et al.*[31] did not find any particular crystallographic difference between the high and low temperature phases that were well refined in the space group *I4/mmm*. The temperature variation of the lattice parameters follows the lattice expansion and no anomaly was observed at $T_{CO}$. Moreover, X-ray absorption spectra (XANES and EXAFS) have not shown any appreciable difference by crossing the transition temperature either.[32] The concluding remark of these studies is the presence of very small distortions in the low temperature phase, below the sensitivity limit for conventional x-ray powder diffraction, neutron powder diffraction and x-ray absorption spectroscopy. Nevertheless, $(1/2,1/2,0)_t$ and $(1/4,1/4,0)_t$ reflections were detected by RXS at the Mn K edge[7]. The observation of a resonance at the $(1/2,1/2,0)_t$ reflections was claimed as an experimental proof for the CO of $Mn^{3+}$ and $Mn^{4+}$ ions following a checkerboard pattern whereas the presence of RXS intensity at the forbidden $(1/4,1/4,0)_t$ reflection was interpreted as due to the *zigzag* OO of the 3*d*- $e_g$ orbitals of $Mn^{3+}$ ions. This interpretation is contrary to several RXS studies on $RE_{0.5}Me_{0.5}MnO_3$ three dimensional manganites which show that the resonances have a structural origin [8-12,15,16]

The purpose of this work is twofold. First, we want to determine and quantify the structural changes present at the phase transition to account for the atomic shifts and their relation with the condensation of phonon modes. Secondly, we will shown, as it was demonstrated for three-dimensional manganites, that RXS scattering at the Mn K edge comes from local structural distortions. For these purposes, we have collected high-resolution synchrotron x-ray powder diffraction patterns for two $La_{1-x}Sr_{1+x}MnO_4$ samples with compositions x=0.5 and x=0.55. Furthermore, we performed RXS measurements at the Mn K edge in a wide set of reflections of a $La_{0.5}Sr_{1.5}MnO_4$ single crystal. The energy, azimuth angle and polarization dependences of these reflections were analyzed. The full information provides us a detailed description of the structural changes at the semiconductor-insulator transition in terms of condensed phonon modes. We conclude that the so-called CO and OO reflections in this layered manganite have their dipolar origin on the local distortions and they are not a proof of ionic charge or d-orbital ordering, as it occurs for other three dimensional mixed-valence transition-metal oxides[33].

## II. EXPERIMENTAL

Polycrystalline $La_{1-x}Sr_{1+x}MnO_4$ (x=0.5 and 0.55) samples were prepared by the ceramic method. Stoichiometric amounts of $La_2O_3$, $SrCO_3$, and $MnCO_3$ were mixed and heated at 1000º C in air overnight. Pellets were sintered at 1450ºC in an oxygen atmosphere for 24 h. This step was repeated three times and the resulting samples were analyzed by x-ray diffraction showing patterns typical of a single phase. These compounds were used for both synchrotron powder x-ray diffraction measurements and single crystal growth.

Crystal growth was made in a homemade floating zone furnace with two semi-elliptical mirrors.[34] The growing was carried out at 2 bars of $O_2$ with a growth speed ranging between 12 and 10 mm per hour. The chemical composition of both pellets and boles, were also tested using the wavelength dispersive x-ray fluorescence spectrometry (Advant'XP+ model from Thermo-ARL). The La:Sr:Mn stoichiometry agreed with the expected values in all cases.

High resolution synchrotron powder x-ray diffraction (XRD) patterns were collected between 80 and 295 K at the ID31 beamline[35] at the European Synchrotron Radiation Facility (ESRF) in Grenoble, France. The sample was loaded in a borosilicate glass capillary ($\phi$=0.5mm) and kept spinning during data acquisition. A short wavelength, $\lambda$=0.3541 Å, was selected to reduce absorption by means of a Ge (1,1,1) double-crystal monochromator. Low temperatures were achieved using an Oxford Cryosystem cryostream equipped with a cold-nitrogen-gas blower. Diffraction patterns were analyzed by the Rietveld method by means of the Fullprof package.[36]

RXS experiments were performed at the ID20 beamline[37] at the ESRF. The incident beam was monochromatized by a Si(1,1,1) double-crystal, energy resolution being 1 eV at the Mn $K$ edge. The beam was linearly polarized (99%) with the polarization vector perpendicular to the scattering plane. The $La_{0.5}Sr_{1.5}MnO_4$ single crystal was cut and polished with the $[1,1,0]_t$ direction as the surface normal. It was mounted inside a closed-cycle helium refrigerator employed for low temperature measurements. The full set up was placed in a four-circles vertical diffractometer. The energy dependence of the scattered signal across the Mn absorption edge was measured, all spectra shown along this paper being corrected for absorption by using the experimental fluorescence. Azimuth scans at the resonance energy were measured by rotating the sample around the scattering vector $Q$, $\varphi$=0º for $k_i$ parallel to $(001)_t$. The polarization of the scattered light was analyzed using a Cu (2,2,0) crystal, which allows the measurement of the σ-σ' and σ-π' channels independently. Theoretical simulations were carried out by the FDMNES code[38].

## III. RESULTS

### A. High resolution synchrotron x-ray powder diffraction

X-ray diffraction measurements were performed on both crushed crystals and polycrystalline seeds. The resulting patterns were very similar for both types of specimens but the powders arising from crushed crystals showed texture effects even after exhaustive grinding. Therefore, these experiments were focused on the second set of samples. At 295 K, both x=0.5 and x=0.55 samples adopt the tetragonal $K_2NiF_4$-structure with space group *I4/mmm*. The x-ray diffraction patterns were analyzed between $2\theta = 2°$ and 38º and the tetragonal phase contributes with a total of 219 reflections in this region. The structural refinement yields atomic positions in perfect agreement with previous publications[31] as can be seen in Table I. Both samples present similar results with similar *c*-axis whereas the *a*-axis is a bit longer for the $La_{0.45}Sr_{1.55}MnO_4$ compound. We also realized the presence of contamination with traces of $La_{1-x}Sr_{2+x}Mn_2O_7$ (~0.4% in *wt*). This impurity has been detected previously[28] and its presence in both crystal and seeds suggest the occurrence of intergrowths. The impurity was also included in the refinements as a secondary phase.

On decreasing temperature, the x-ray patterns resemble each other. The only noticeable change is a small splitting of some diffraction peaks (inset of Fig. 1) showing that the low temperature phase is no longer tetragonal but orthorhombic. We were unsuccessful in detecting $(h/2,h/2,l)_t$ or $(h/4,h/4,l)_t$ (*h* odd, *l* even) superstructure reflections indicating that their intensities are too weak to be detected even by synchrotron powder XRD. The low temperature phase can be refined under an orthorhombic cell whose unit cell is metrically related to the tetragonal one as $c_t \times \sqrt{2}\, a_t \times \sqrt{2}\, a_t$ ($a_t$ and $c_t$ being the tetragonal lattice parameters). The resulting space group is *Cmmm* and the tetragonal *ab*-plane corresponds to the *bc*-plane in the orthorhombic cell. The structural refinement for $La_{0.45}Sr_{1.55}MnO_4$ is plotted in Fig. 1 and the refined data for both samples are also included in Table I. These data show that fractional coordinates lies close to the theoretical tetragonal positions and in some cases within the standard deviation. This structure was already reported by Sternlieb *et al.*[28] There are two different crystallographic sites for Mn and three different sites for O forming an octahedron around each Mn (Table I). The O11 and O12 with the same *x*-coordinate as Mn atoms, equal *bc*-plane, represent the equatorial oxygens of the octahedra while O21 and O22 are the apical oxygens for Mn1 and Mn2, respectively. Finally, Mn1 and Mn2 environments compose a checkerboard pattern in the *bc*-plane.

The temperature dependence of the lattice constants for x=0.5 and x=0.55 samples are shown in Figure 2. The behavior of these compounds is nearly identical and they show a similar small orthorhombic distortion at low temperature but the difference between the Mn-O equatorial distances for the two Mn sites is insignificant. The phase transition occurs between 220 and 240 K in agreement with the reported $T_{CO}$ for both compositions.[39] Previous publications and our RXS measurements (see below) report the unambiguously observed $(h/2,h/2,l)_t$ and resonant $(h/4,h/4,l)_t$ (*h* odd, *l* even in both cases) superstructure reflections. The *Cmmm* cell can account for $(h/2,h/2,l)_t$ reflections but not for the $(h/4,h/4,l)_t$ ones. The appearing of the latter reflections requires the additional doubling of one of the shorter orthorhombic axes.

In order to explore the possible atomic displacements able to result in the mentioned orthorhombic cell, we have used the ISODISPLACE tool.[40] This software allows exploring structural phase transitions generating atomic displacement patterns induced by irreducible representations of the parent space-group symmetry. The tetragonal high temperature phase was used as the parent compound and we have searched over arbitrary k-points of the first Brillouin zone in order to obtain orthorhombic cells with $2\sqrt{2}\ a_t \times \sqrt{2}\ a_t \times c_t$ lattice parameters. Among the different solutions, only the C-centered cells were retained as the reflection extinctions due to the C-lattice condition is observed in the full pattern. There are eight possible solutions with the point group *mmm* listed in the Table II. In order to select the most appropriate model, we have made use of our RXS results. Solutions 3, 4 and 8 can be excluded as they only permit a single crystallographic site for Mn atoms with the same point group and measured RXS signals for $(h/2,h/2,l)_t$ reflections, coming from any charge segregation, would not be possible. We have observed $(9/4,7/4,0)_t$ and $(7/4,9/4,0)_t$ superstructure reflections with significant Thomson scattering, i.e. these reflections are allowed. According to the respective transformation matrices, these reflections corresponds to the orthorhombic (4,0,1)+ (1/2,0,8) reflections for the setting of solution 1 and the (0,4,1)+(0,1/2,8) reflections for the rest of solutions (2, 5, 6 and 7). These reflections are forbidden for solutions 1 and 6 so they can be discarded. We have also observed that $(7/4,7/4,0)_t$, $(7/4,7/4,2)_t$ and $(9/4,9/4,2)_t$ only show resonant scattering without Thomson contribution so the three reflections are forbidden by symmetry. They correspond to the (0,0,7)+(0,7/2,0), (2,0,7)+(2,7/2,0) and (2,0,9)+(2,9/2,0) orthorhombic reflections, respectively. The only space group where these reflections are forbidden corresponds to the solution 2, which concurs with the space group initially used by Larochelle *et al.*[29] in their single crystal diffraction experiment. However, these authors have observed allowed Bragg reflections with wavevector $(1/4,1/4,l)_t$ suggesting a lowering of symmetry from *Bbmm* to *B2mm*. We did not observe Thomson scattering for this type of reflections in agreement with the results reported by either Sternlieb et al.[28] or Murakami et al.[7]. We think this controversy might arise from the existence of multiple reflections (or Renninger scattering) with strong contribution around $(1/4\ 1/4\ l)_t$ reflections.

The symmetry analysis is compatible with a direct transition from the parent structure *I4/mmm* to the orthorhombic *Cmcm*. The atomic coordinates of the undistorted *Cmcm* structure are also listed in the Table III. There are three active modes capable of producing atomic displacements: $\Gamma_1^+$, $\Delta_2$ and $X_1^+$. The atomic displacements associated to each condensed individual mode are collected in the Table IV. The primary mode should be $\Delta_2$ acting on the $(1/4,1/4,0)_t$ k-point and it could be connected to the appearing of the so-called OO reflections. Another important mode is the $X_1^+$ associated to the $(1/2,1/2,0)_t$ k-point which allows the appearing of the so-called CO reflections. $\Gamma_1^+$ coupled to the $(0,0,0)_t$ k-point is ascribed to the expansion (or contraction) of some bond lengths (rock-salt layer) along the *x*-direction.

Only one of the three modes, $\Delta_2$, affects to the Mn sites. There are two crystallographic sites for Mn atoms: Mn1 is located at the 4*a* Wyckoff position with a point group 2/*m*.. while Mn2 is placed at 4*c* with *m2m* point group. $\Delta_2$ only acts on Mn2 and the movement is along the orthorhombic *y*-direction. This mode produces similar motions to the Sr/La2 and O12 atoms. All of the apical oxygen atoms and Sr/La atoms could be affected by the condensation of a $X_1^+$ mode which is related to atomic displacements along the *x*-direction. In opposite to the effect induced by $\Gamma_1^+$, the shifts produced by the $X_1^+$ mode are in opposite directions for each couple of atoms. In the case of apical oxygens, the condensation of this mode leads to an ordered array of large and short Mn-O$_{apical}$ distances which gives rise to a checkerboard pattern of expanded and contracted (along the *x*-axis) MnO$_6$ octahedra in the *bc*-plane.

The only condensed modes able to produce displacements along the *z*-direction (i.e. responsible of doubling the *c*-axis) are two $\Delta_2$ and two $X_1^+$ modes acting on the equatorial oxygen atoms (named as O11 and O12 in Table III). They would have a main role in the appearing of the OO superstructure peaks. Figure 3 shows the displacement of the equatorial O atoms associated to these $\Delta_2$ and $X_1^+$ modes. The $\Delta_2$(B2u) mode is an asymmetric stretching mode and the corresponding deformation is similar to the one expected from a J-T distortion where two O approach one Mn site while they move away from the other one. $X_1^+$(B2u) is a stretching mode (breathing-type) where the Mn-O bonds shorten for one Mn site while they enlarge for the neighbor Mn-site as indicated in Fig. 3. Both modes are presumed to be the driving force for the CO transition according to previous works.[23] $\Delta_2$(B3u) is a bending mode associated to the tilt of equatorial oxygen atoms around the Mn1 in opposite directions what implies the coupled enlargement and shortening of the O-Mn2-O angles. $X_1^+$(B3u) is also a bending mode affecting both O-Mn1-O and O-Mn2-O bond angles in a way that resembles the movement of the scissors blades rotating around the Mn2 site. In the resonant study that follows we consider the role of all modes in order to clarify the previous conjecture about the CO transition.

There is no reason to refine the XRD patterns using this structure since the technique is not sensitive to the superstructure peaks. However, in order to take full advantage of the symmetry analysis, we have performed simulations of the expected patterns for different amplitudes of the condensing modes. We have used the AMPLIMODES program[41] to decompose the symmetry-breaking distortion present in the transition from *I4/mmm* to *Cmcm* (solution 2) into contributions from different symmetry-adapted modes. We have tested different amplitudes of the modes to find the lowest limit to detect noticeable superstructure peaks in the pattern. This limit is then considered as the maximum distortion compatible with our experimental results. For example, our simulations indicate that the maximum oxygen displacement of any of the modes indicated in the Fig. 3 should be shorter than 0.09 Å because longer shifts would produce noticeable superstructure peaks by synchrotron XRD. The atomic displacements of Mn2 or La/Sr atoms must be even much smaller, their respective sensitivity limits ranging between 0.04 and 0.01 Å.

**B. Resonant x-ray scattering**

## B1. Experimental results

The dependence of the scattered intensity as a function of the incoming photon energy for the reflections $(3/2,3/2,0)_t$ and $(5/2,5/2,0)_t$ across the Mn K edge in the σ-σ' channel at different azimuth angles ($\varphi$) is shown in Figure 4. We observe scattered intensity at energies below the Mn K absorption edge, which comes from the atomic displacements out of the *I4/mmm* symmetry (Thomson scattering). Therefore, these reflections are permitted by the crystal symmetry of the low temperature phase. On the other hand, a strong enhancement of the intensity, a broad main resonance, is observed for the $(3/2,3/2,0)_t$ reflection at photon energies close to the Mn K edge. This enhancement comes from the difference between the anomalous atomic scattering factors of Mn atoms located in different crystallographic sites. Contrary to $(3/2,3/2,0)_t$, the $(5/2,5/2,0)_t$ reflection shows a strong decrease of the intensity at the Mn K edge. A similar change in the sign of the resonance has been observed in the RXS study of layered cobaltites and ferrites.[42,43] The opposite behavior of the energy dependent spectra for these two reflections indicates that the anomalous part of the atomic scattering factor is in phase with the Thomson term for the $(3/2,3/2,0)_t$ reflection and out-of phase for the $(5/2,5/2,0)_t$. The spectral shape evolves with the azimuth angle and the overall intensity at the maxima shows moderate marked azimuth dependence. This last effect is the consequence of an anisotropy in the Mn anomalous atomic scattering factors.

Figure 5 shows the spectra of the $(5/2,5/2,0)_t$ reflection in the σ-π' channel at different azimuth angles and the $(3/2,3/2,0)_t$ at $\varphi=45$ as a function of the incoming photon energy. The intensity of these reflections was corrected from the σ−σ´ contamination coming from the non-exactly 90º reflection of the Cu(220) analyzer. These reflections show a double peak structure and strong azimuth dependence with $\pi/2$ period. The inset of Fig.5 shows the latter at the main resonance (E=6.557 keV). We note that the shape of the energy dependent spectra does not change either with the azimuth or with the scattering vector. The observation of a signal in this channel is also a signature of the anisotropy of the anomalous atomic scattering factors of the individual Mn atoms as we will explain in Section B2.

Reflections of the $(h/4,h/4,0)_t$ -type with *h*=odd corresponding to the so-called OO were also investigated. No scattered intensity was observed for these reflections at any energy of the photons in the σ-σ' channel but a strong resonance at the photon energy of the Mn K edge is observed in the σ-π' channel. The lack of a σ-σ' signal demonstrates that these reflections are forbidden by symmetry and they can be classified as ATS reflections[44-46] Figure 6 shows the energy dependence spectra of the $(h/4,h/4,0)_t$ ($h$=odd) reflection at $\varphi$= 0° and consecutive $h$ indices (5,7,9). The shape of the energy spectrum is the same regardless of the azimuth (not shown) and the modulus of the ***Q*** vector, as expected for dipolar 1$s$ to 4$p$ electronic transitions. Only at the pre-edge region we realize the presence of little differences, probably due to small hybrid dipole-quadrupole and/or pure quadrupolar contributions. The azimuth dependence has a π period as shown in the inset of Figure 6. On the other hand, the intensities at different ***Q*** follow a $\cos^2\theta$ dependence. In order to check the origin for $(1/4,1/4,0)_t$ period, we have also plotted in Fig. 6 the energy dependence of the $(9/4,7/4,0)_t$ reflection recorded in the σ-σ' channel. The observation of this reflection in the unrotated polarization channel beyond and below the absorption edge (Thomson scattering) indicates that the $(h/4,h/4,0)_t$ periodicity is structural.

The intensity of both type of reflections, $(h/2,h/2,0)_t$ and $(h/4,h/4,0)_t$, decreases with increasing temperature. For instance, the temperature dependence of $(3/2,3/2,0)_t$ and $(9/4,9/4,0)_t$ is plotted in the Fig. 7. These reflections vanish at $T_{CO}$~230 K which agrees with the suppression of the orthorhombic distortion (see Fig. 2) as well as with the change in the electrical resistivity. The temperature evolution follows a similar trend at low temperature for both reflections but it shows noticeable differences close to $T_{CO}$. The intensity of the $(3/2,3/2,0)_t$ reflection monotonously decreases on warming while $(9/4,9/4,0)_t$ suffers little evolution between 140 and 190 K and then, a more drastic decay is observed around 200K. As usually found in half-doped manganites,[10,12,30] the modulation wave vector remains constant and commensurate with the lattice in the whole temperature range. This result however contrasts with the temperature evolution of the modulation wave vector observed by electron diffraction in some nearly half-doped manganites[47].

**B2. Theoretical framework of the resonant x-ray scattering**

As it was concluded in Section 2, we assume that the low temperature symmetry is *Cmcm* (standard group n° 65) as also proposed by Larochelle et al.[29,30] and we will use this description to interpret the RXS results. The orthorhombic lattice parameters are $a_o$=12.398382 Å, $b_o$= 5.444881 Å, $c_o$= 10.895792 Å, which are related to the tetragonal *I4/mmm* setting by $\vec{a}_0 = \vec{c}_t$, $\vec{b}_0 = \vec{a}_t - \vec{b}_t$ and $\vec{c}_0 = 2(\vec{a}_t + \vec{b}_t)$. The $(h/2,h/2,0)_t$ and $(h/2,-h/2,0)_t$ with $h$=odd reflections correspond to the $(0,0,2h)_o$ and $(0,h,0)_o$ reflections in the orthorhombic *Cmcm*, respectively. The $(0,h,0)_o$ reflections are extinct in this symmetry. Therefore, no mixture between these two reflections due to the twinning is expected. The $(h/4,h/4,l)_t$ with $h$=odd and $l$=even reflections correspond to the $(l, 0, h)$ ones in *Cmcm*. We first assume that the two different Mn sites are in special positions, Mn1 in (0,0,0) and Mn2 in (0,0.5,0.75). Hereafter, we will refer to the orthorhombic structure.

The structure factor for $(0,0,2h)_o$ ($h$=odd) reflections is given by $F = F_T + 4(f_{Mn1} - f_{Mn2})$, where $F_T$ is the Thomson structure factor coming from atomic motion and $f_{Mni}$ ($i$=1,2) is the anomalous atomic scattering factor for each of the Mn atoms. According to the *z*-coordinates of the atoms (Tables III, IV), the Thomson contribution from Mn, Sr/La and apical oxygens is null so $F_T$ can only arise from motions of the equatorial oxygens. The $(l,0,h)_o$ Thomson reflections with $h$=odd and $l$=even are systematically extinct by the presence of the c-glide within the orthorhombic *ac*-plane. They are ATS reflections and can be permitted by the anisotropy of the anomalous atomic scattering tensor in equivalent Mn atoms. We use the dipole approximation to describe the anomalous atomic scattering tensor[44-46]. This approach is fully justified at the absorption K edge because spectroscopic features arising from higher order transitions are generally observed at the pre-edge. Within this approach, the atomic scattering factor can be expressed as $f_{\varepsilon,\varepsilon'} = \varepsilon' \widehat{f} \varepsilon$. Here, $\varepsilon$ and $\varepsilon'$ are the incident and scattered electric polarization vectors and $\widehat{f}$ is a second rank symmetric tensor which characterizes the sample properties at a given energy, and it is independent of the radiation. The specific form of the scattering tensor depends on the point symmetry of the scattered atom. Therefore, applying the symmetry operations for the two Mn sites we can obtain the general form of the scattering tensor. The Mn1 site has *2/m..* symmetry, i.e. the associated tensor is invariant under a 180º rotation around the *a*-axis and the mirror *bc*-plane. Thus,

$$\widehat{f}(Mn1) = \begin{pmatrix} f_{xx}^1 & 0 & 0 \\ 0 & f_{yy}^1 & f_{yz}^1 \\ 0 & f_{zy}^1 & f_{zz}^1 \end{pmatrix}$$

For the Mn2 atom, the symmetry is *m2m*, 180º rotation around *b*-axis and mirror *bc*- and *ab*-planes. In this case,

$$\hat{f}(Mn2) = \begin{pmatrix} f_{xx}^2 & 0 & 0 \\ 0 & f_{yy}^2 & 0 \\ 0 & 0 & f_{zz}^2 \end{pmatrix}$$

Taking into account that the Mn atoms are correlated by the *ac*-glide plane along the c direction, $f$(Mn2) = $f$(Mn2') and the out-of-diagonal $f_{yz}$ component in $f$(Mn1) changes sign when going from Mn1 to Mn1'

Applying these conditions, the anomalous scattering structure factor for the $(0,0,2h)_{o,h=odd}$ and $(l,0,h)_{o,h=odd,\ l=even}$ reflections in the σ-σ' and σ-π' channels are (the superscripts on $f$ refer to the Mn site number):

$$F_{Mn}(0,0,2h)_{\sigma\sigma'} = 4\left[(f_{xx}^1 - f_{xx}^2)\sin^2\varphi + (f_{yy}^1 - f_{yy}^2)\cos^2\varphi\right] \quad (1)$$

$$F(0,0,2h)_{\sigma\pi'} = 4\sin\varphi\cos\varphi\sin\theta\left[(f_{xx}^1 - f_{yy}^1) - (f_{xx}^2 - f_{yy}^2)\right] \quad (2)$$

$$F(l,0,h)_{\sigma\pi'} = 4\left[f_{yz}^1 \cos\theta\sin\varphi\right] \quad (3)$$

We note here that a Thomson scattering term must be added to the expression (1) for the $(0,0,2h)_{o,h=odd}$ reflections to get the total structure factor in the σ-σ' channel. The contribution from an oxygen motion Δz (unit cell distances) to the structure factor of the $(0,0,2h)_{o,h=odd}$ reflections is given by: $F_O(0,0,2h) = f_O (-1)^{n+1}(4n+2)16\pi\Delta z$, with $2h=4n+2$. We observe that this contribution to the structure factor changes sign with n whereas the real part of $\Delta f_{xx} = (f_{xx}^1 - f_{xx}^2)$ and $\Delta f_{yy} = (f_{yy}^1 - f_{yy}^2)$ maintain their sign at energies close to the main resonance and they are positive for all $(0,0,2h)_{o,h=odd}$ reflections. Because this resonant term and the Thomson scattering are in phase for the $(0,0,6)_o$ reflection, a peak in the energy dependence of this reflection is observed at the Mn K edge (Fig. 4). On the other hand, the two terms subtract for the $(0,0,10)_o$ reflection and a valley is observed. Moreover, the imaginary part of $\Delta f_{xx}$ and $\Delta f_{yy}$ are also > 0 at the absorption threshold. This implies that the atomic anomalous scattering factor for Mn1 atom is greater than the one for Mn2 at the resonance, *i.e.* the polarized x-ray absorption spectrum for the Mn1 atom is at lower energies than the one for Mn2 in the *x* and *y* directions, which is in agreement with the coupled expansion of the Mn1O$_6$ and compression of the Mn2O$_6$ octahedra given by the oxygen motion Δz.

We observed azimuthal dependence of these reflections, which indicates that the atomic scattering factors for both Mn1 and Mn2 are anisotropic. We also observe σ-π' intensity for $(0,0,2h)_{o,h=odd}$ reflections, which confirms the presence of this anisotropy. As it can be deduced from Eq. (2), the occurrence of these reflections in the σ-π' channel is due to the fact that $(f_{xx}^1 - f_{yy}^1)$ and/or $(f_{xx}^2 - f_{yy}^2)$ are different from zero and the period π/2 is followed.

Regarding the forbidden $(0,0,h)_{o,h=odd}$ reflections, the azimuthal, polarization and ***Q*** dependence are nicely described by equation (3). We observe that the shape of the energy dependent RXS spectra does not change with either the azimuth or the ***Q*** scattering vector, which implies that it is only determined by the $f_{yz}$ component of Mn1 atoms. This nonzero out–of-diagonal component comes from a distortion of the Mn1O$_6$ octahedron in the *bc*-plane, one Mn-O bond is enlarged and the other is shortened.[9]

### B3. Simulations and analysis of the resonant x-ray scattering

In order to get a reliable comparison between the experimental intensities and the theoretical simulations we have estimated the intensity of the so-called CO and OO reflections here studied by comparison with Thomson intense Bragg reflections within a relative error of about 30%. We have followed two approaches to describe the energy dependence of the studied reflections: the semiempirical approach[7,10,12,14,16] and the simulations performed using the FDMNES code.

The semi-empirical procedure makes use of the experimental polarized x-ray absorption spectra on the same single-crystals.[32,48] Assuming that the difference between the components of the atomic scattering tensor comes exclusively from the different energy position of the absorption edge, the only fitting parameter is the chemical shift which could be correlated with the difference in the formal valence state between both Mn atoms.[49,50] We note that the x-ray absorption coefficients with the polarization vector along either the *x* or *y* directions are $\mu_{xx}(\exp)=(\mu_{xx}(Mn1)+\mu_{xx}(Mn2))/2$ and $\mu_{yy}(\exp)=(\mu_{yy}(Mn1)+\mu_{yy}(Mn2))/2$, respectively. We assume that $\mu_{xx}(Mn1)(E)=\mu_{xx}(\exp)(E-\delta E)$ and $\mu_{xx}(Mn2)(E)=\mu_{xx}(\exp)(E+\delta E)$ and the same approach applies to the *y* component. We remind here that 2$\delta$ is referred in the literature as the chemical shift.[49] As it is well known, the anomalous atomic scattering factor is given by $f'(E)+if''(E)$, where the imaginary part is related to the x-ray absorption coefficient $\mu(E)$ through the optical theorem by $f''(E)=(mcE/2e^2h)\mu(E)$. The real part $f'(E)$ is in turn related to the imaginary one $f''(E)$ through the mutual Kramers-Kronig relation. Using these relations, we have calculated the $f_{xx}$ and $f_{yy}$ components for the two Mn atoms.

The experimental RXS spectra of the $(0,0,2h)_{o,h=odd}$ reflections in the σ-σ' channel were fitted to the following expression

$$I(0,0,2h)_{\sigma\sigma'} = \left[ 4\left[ (f_{xx}^1 - f_{xx}^2)\sin^2\varphi + (f_{yy}^1 - f_{yy}^2)\cos^2\varphi \right] \pm f_{thomson} \right]^2 \quad (5)$$

Only, the chemical shift $2\delta$ was fitted and the Thomson term was fixed to the experimental non-resonant intensity below the absorption edge. The comparison between the best fits and the experimental spectra is quite good as it can be observed in Figure 8(a) and 8(b) for the $(0,0,6)_o$ and $(0,0,10)_o$ reflections, respectively. The obtained $2\delta$ values are 0.3 eV in the $x$ direction and 0.7 eV in the $y$ direction. Using the same parameters, the energy dependence of the RXS intensity in the σ-π' channel for these two reflections was also nicely reproduced by the square of equation (2), as it is shown for the $(0,0,10)_o$ reflection in Fig. 8(c). The chemical shift can be correlated with the formal valence of the Mn atom, assuming a linear relationship. Taking into account that the chemical shift between formal $Mn^{3+}$ and $Mn^{4+}$ atoms is about 4.5 eV[49,50], the estimated charge disproportionation between Mn1 and Mn2 would be about 0.15 electrons.

In order to simulate the anisotropy of the Mn1 atom, responsible for the appearance of $(0,0,h)_{o,h=odd}$ forbidden reflections, we approximate that the $f_{yz}$ term comes from the difference between the $f_{yy}$ and $f_{zz}$ diagonal terms of the Mn1 atomic scattering tensor when the reference frame is along the Mn-O bonds. This approach was used in previous works using a model with a tetragonal distortion.[10,12,16] Similarly as for $(0,0,2h)_{o,h=odd}$, this anisotropy is simulated through an energy shift between the polarized x-ray absorption coefficients and consequently, between the atomic scattering factors. In this way, $f_{yz}(E)=f_{yy}(E+\delta')-f_{zz}(E-\delta')$, $2\delta'$ being the anisotropic energy shift[9]. The RXS intensity of these forbidden reflections would then be expressed as:

$$I(0,0,h)_{\sigma\pi'} = 16\left[ (f_{yy}(E+\delta') - f_{zz}(E-\delta'))\cos\theta\sin\varphi\cos\varphi \right]^2. \quad (6)$$

The best fit obtained using this expression is shown in Figure 8d for the $(0,0,7)_o$ reflection, for $2\delta'=0.9$ eV. We observe that the simulation is slightly poorer than previous ones since the experimental $f_{yy}$ and $f_{zz}$ terms also include the Mn2 atomic scattering factor. However, it reproduces quite well the absolute intensity of the resonance showing that resonances in ATS reflections are mainly originated by an energy shift between the anisotropic components of the anomalous atomic scattering factor. The obtained anisotropic energy shift correlates well with those found for several half-doped manganites thin films that are also of the order of 1 eV,[16] but it is a bit smaller than the deduced for the three-dimensional bulk half-doped manganites, around 1.5 eV.[10,12] This anisotropic energy shift will give us an estimation of the magnitude of the tetragonal distortion for the Mn1O$_6$ octahedron, which results to be about 0.05 Å.[16]

As we have already said, the RXS experimental spectra were also analyzed using the FDMNES code.[38] This program calculates from first principles, the x-ray absorption coefficient and the anomalous atomic scattering factor tensor as a function of the photon energy for the different crystallographic Mn sites. The intensity of resonant reflections is then calculated using these atomic scattering factors for a finite cluster of atoms surrounding the absorbing one and the potential used is made with the neutral atomic spheres. This approach qualitatively agrees with the experiment and has allowed us to determine the effect of the geometry on the anomalous atomic scattering factor. However, the code is not implemented to calculate the scattering from a random atomic distribution as present in our sample, *i.e.* La and Sr atoms occupy the same crystallographic site in a relative weight given by the composition. In order to overcome this problem, we have calculated the atomic scattering factor tensors for clusters with either only Sr or only La atoms. Final atomic scattering factors are obtained as the weighted (3 to 1) addition of these independent calculations.

We used the *Cmcm* space group and the different active modes for the oxygen atoms given in Table IV were checked independently. The $\Delta_2$(B3u) mode does not produce any resonance either for the $(0,0,2h)_{o,h=odd}$ or for the $(0,0,h)_{o,h=odd}$ reflections. The $\Delta_2$(B2u) mode corresponds to a pseudo J-T distortion of Mn1 atoms in only the *bc*-plane and it is the responsible for the occurrence of the $(0,0,h)_{o,h=odd}$ forbidden reflections. This mode does not induce Thomson scattering in $(0,0,2h)_{o,h=odd}$ reflections. The $X_1^+$(B2u) and $X_1^+$(B3u) modes can explain the appearance of Thomson scattering and the resonance for $\varphi=0°$ in $(0,0,2h)_{o,h=odd}$ reflections. $X_1^+$(B2u) is a breathing mode, the Mn(1)O$_4$ square expands and the Mn(2)O$_4$ one contracts whereas in $X_1^+$(B3u), O atoms move along *z* oppositely to the movement along *y*. Finally, $X_1^+$(A1) mode with oxygen motion along the $a_o$-axis is responsible of the resonant scattering of the $(0,0,2h)_{o,h=odd}$ reflections for $\varphi=90°$. In order to discriminate which of the two modes, $X_1^+$(B2u) or $X_1^+$(B3u), is active, we carried out theoretical simulations using FDMNES. They allow us to conclude that we only need to add a compression (expansion) along the $c_o$-axis to the J-T $\Delta_2$(B2u) and apical $X_1$ modes to reproduce the overall RXS energy and its azimuth dependence. In fact the simulations with either only the $X_1^+$(B2u) breathing mode or with a linear combination of the two modes —$X_1^+$(B2u)+ $X_1^+$(B3u) — give similar results. We also investigated the effect of the coupled *zigzag* displacement of Mn2 atoms given by the $\Delta_2$(B2u) mode along the *y* direction, but the RXS simulations demonstrated a very limited sensitivity to this movement. Therefore, we end up only with three active modes: the *J-T* $\Delta_2$(B2u), breathing $X_1^+$(B2u) and apical $X_1^+$(A1) to fit RXS spectra using FDMNES. The refined parameters were $\delta_1$, $\delta_3$ and $\delta_4$, as defined in Table IV.

Figure 9 shows the comparison between experimental spectra and the best fit calculations The calculated polarized XANES spectra nicely agree with the experimental ones[32] as it is shown in the upper panel of Fig. 9. The polarized XANES spectra show a macroscopic anisotropy (the spectrum with $\varepsilon$ along the (1,0,0) direction is shifted at lower energies). The simulations are good regarding the overall behavior of the reflections but they cannot be considered as quantitative. The main fit parameter was the absolute and relative (resonant to non-resonant) intensities. We obtained $\delta_1$ = 0.0025(4) r.l.u ; $\delta_3$ = 0.0011(4) r.l.u. and $\delta_4$ = 0.0004(2) r.l.u (r.l.u. = reciprocal lattice units). The values of the refined atomic displacements are too low to be detected by high resolution powder XRD. Indeed, absolute resonant intensities are under ~$10^{-4}$ times that of an intense Bragg allowed reflection. It is noteworthy that although the amplitude of the $\Delta_2$(B2u) mode is the largest one, the other two modes are necessary to reproduce the RXS spectra for $(0,0,2h)_{o,h=odd}$ reflections, including the different behavior of $(0,0,6)_o$ and $(0,0,10)_o$ reflections and the appearance of resonances at $\varphi=90°$. Figure 10 outlines the local structure around the two Mn atoms including the values of the Mn-O bond lengths as extracted from this study. It is clear that both MnO$_6$ octahedra are distorted and the largest distance corresponds to the apical O atoms in both cases, i.e. the out of *bc*-plane direction (orthorhombic description).

## VI. DISCUSSION AND CONCLUSIONS

High resolution x-ray powder diffraction reveals a structural phase transition at the semiconductor-insulator transition temperature. A detailed symmetry-mode analysis of the transitions from a tetragonal cell (*I4/mmm*) to an orthorhombic one with *mmm* point group reveals that the *Cmcm* space group is the only model to account for our RXS experiments at the Mn K edge. As it was previously reported,[7,28-31,51] the low temperature phase is well described by a checkerboard ordering of two inequivalent Mn sites (Mn1 and Mn2). Superstructure peaks are hardly noticeable because the metal atoms (La/Sr and Mn) practically remain in the undistorted tetragonal positions. Hence, the structural transition from *I4/mmm* to *Cmcm* is driven by the motion of the oxygen atoms in accordance to three active modes. There are two symmetric stretching modes —the equatorial $X_1^+$(B2u) and the apical $X_1^+$(A1) — and one asymmetric stretching mode, $\Delta_2$(B2u), which produces the anisotropy on the Mn1 sites. We note that this description is similar to previous proposals but a significant point must be emphasized. In previous models, only the distortions arising from the $\Delta_2$(B2u) were taken into account in some studies[29,51] while other authors[7,28] only considered distortions arising from the $X_1^+$(B2u) mode. Our work reveals that all of the three modes are necessary to properly describe the RXS experimental data regarding superstructure $(h/2,h/2,0)_t$ reflections.

The most striking point of this study, only softly outlined previously,[29] is that the two inequivalent Mn cannot be ascribed to $Mn^{3+}$ and $Mn^{4+}$ ions. The analysis of the RXS spectra using a semi-empirical phenomenological model has shown that the chemical shift found between the anomalous atomic scattering factors of the two Mn sites at the Mn K edge is markedly smaller than the experimental chemical shift measured between formal $Mn^{3+}$ and $Mn^{4+}$ reference compounds. From our measurements, we estimate the charge disproportionation between both sites to be rather modest, about 0.15 electrons. Theoretical calculations carried out with the FDMNES program have allowed us to determine the atomic displacements for the oxygen atoms compatible with RXS data, which turn out to be below 0.027 Å for the most intense mode. The Bond Valence Sum method[52] yields values of +3.5 and +3.8 valences for Mn1 and Mn2 atoms, respectively. The disproportionation obtained by this method coincides with the estimated previously in the semi-empirical model. We remind here that the local geometry and the XANES chemical shift are intimately correlated[50] being the determination of the valence exclusively formal. In fact, the real Mn 3*d* occupation in manganites is not yet a fully resolved issue. Recently, a study on $La_{1-x}Sr_{1+x}MnO_4$ samples using the Mn K$_\beta$ emission spectroscopy has revealed that hole doping causes a spatial redistribution of the Mn 3*d* electrons without reducing the total charge on Mn atoms, the doped holes mainly have O *2p* character.[48] This description shares the theoretical results of Ferrari et al.,[53] who proposed that the CO-OO ordering in La$_{0.5}$Ca$_{0.5}$MnO$_3$ is better described by a charge-density wave of oxygen holes. Not far from this vision, Okuyama *et al.*[54] more recently reported on a possible correlation between the dimensionality of the lattice and the charge segregation in manganites, which they found to be far from ionic in all studied cases by an x-ray structural study. They conclude that single-layered (*n*=1) systems should generally show a negligible charge disproportionation between Mn sites, notably smaller than for the non-layered case (*n*=∞). However, our present study yields a value that is comparable to that found in other three dimensional manganites[10].

Another important point concerns the interpretation of the ($h/4,h/4,0$)$_t$ (*h*=odd) forbidden reflections. Their origin was explained in terms of an OO of Mn $e_g$ states. Instead, local structural distortions have also been proposed to be the main responsible for the observed anisotropy in manganites[9,15]. This work clearly confirms that the resonant component of these reflections (the nonresonant part is extinct by the crystal symmetry), permitted by the presence of a *c*-glide plane in the space group *Cmcm*, arises from the local geometrical asymmetry of the Mn1 atoms in a way that resembles previous RXS studies on LaMnO$_3$.[13] As it will be shown in Fig. 10, the distortion around the Mn1 atom (conventionally assigned to a J-T tetragonal distorted Mn$^{3+}$) is not tetragonal but orthorhombic, corresponding the longest Mn-O distance to the apical oxygen atom. On the other hand, the Mn2 atom (generally assigned to an isotropic Mn$^{4+}$) shows a similar degree of distortion and the longest Mn-O distance corresponding to the apical one. This fact explains the macroscopic anisotropy observed in the polarized XANES spectra (Fig. 9). Therefore, both the small charge disproportion between the two Mn atoms and the minor (not tetragonal) distortion on the Mn1 atoms are inconsistent with the previous assignment of Mn1 to a J-T tetragonal distorted Mn$^{3+}$. Hence, the associated ordering of the Mn 3*d* $e_g$ orbital has no physical meaning. This implies that the interpretation of the soft RXS spectra at the Mn L$_{2,3}$ edge as an OO of $e_g$ electrons.[17-18,20-23] must be revised in order to make it coherent with the structural results.

In summary, the semiconductor-insulator phase transition in La$_{0.5}$Sr$_{1.5}$MnO$_4$ is well described as originated in the associated structural changes, which lower the crystal symmetry from the tetragonal *I4/mmm* to the orthorhombic *Cmcm* phase. Three soft modes would drive the process: first, $\Delta_2$(B2u), associated to a pseudo Jahn-Teller distortion in the *bc*-plane, and second, with a smaller amplitude, an overall contraction of Mn2 atoms and expansion of Mn1 ones ($X_1^+$(B2u) and $X_1^+$(A1)). Although the three modes condense at the same temperature, their temperature dependence is different, probably related to the hierarchy of the distortions driven the phase transition. The condensation of these modes establishes a checkerboard ordering of two different Mn sites, whose valence states turn out to be very similar. On the other hand, the local geometrical anisotropy and ordering of Mn1 atoms is due to the $\Delta_2$(B2u) mode.

## ACKNOWLEDGEMENTS

The authors thank ESRF for granting beam time and ID31 beamline staff for their assistance during the experiments. We also thank Y. Joly for providing the FDMNES code and for fruitful scientific discussions. This work has been financially supported by the Spanish MICINN (project FIS08-03951) and D.G.A. (CAMRADS project).

|  | La$_{0.45}$Sr$_{1.55}$MnO$_4$ | | La$_{0.5}$Sr$_{1.5}$MnO$_4$ | |
|---|---|---|---|---|
| **T (K)** | 295 | 80 | 295 | 80 |
| **Space group** | *I4/mmm* | *Cmmm* | *I4/mmm* | *Cmmm* |
| **a (Å)** | 3.85908(1) | 12.39816(2) | 3.85850(1) | 12.39902(2) |
| **b (Å)** | - | 5.44479(4) | - | 5.44418(3) |
| **c (Å)** | 12.41490(2) | 5.44782(4) | 12.41537(3) | 5.44711(3) |
| **Mn1:** | *(2a) 000* | *(2a) 000* | *(2a) 000* | *(2a) 000* |
| **B (Å$^2$)** | 0.29(1) | 0.29(13) | 0.32(2) | 0.22(9) |
| **Mn2:** | - | *(2c) 0½½* | - | *(2c) 0½½* |
| **B (Å$^2$)** | - | 0.21(13) | - | 0.13(8) |
| **Sr/La1:** | *(4e) 00z* | *(4g) x00* | *(4e) 00z* | *(4g) x00* |
| **z or x** | 0.35789(1) | 0.1419(2) | 0.35779(3) | 0.1421(2) |
| **B (Å$^2$)** | 0.42(1) | 0.19(4) | 0.55(1) | 0.34(3) |
| **Sr/La2:** | - | *(4h) x½½* | - | *(4h) x½½* |
| **x** | - | 0.3579(2) | - | 0.3579(2) |
| **B (Å$^2$)** | - | 0.28(4) | - | 0.34(3) |
| **O1:** | *(4c) 0½0* | *(8n) 0yz* | *(4c) 0½0* | *(8n) 0yz* |
| **y** | - | 0.2519(37) | - | 0.2577(18) |
| **z** | - | 0.2502(35) | - | 0.2434(17) |
| **B (Å$^2$)** | 0.71(3) | 0.65(3) | 0.95(6) | 0.46(3) |
| **O21:** | *(4e) 00z* | *(4g) x00* | *(4e) 00z* | *(4g) x00* |
| **z or x** | 0.1606(1) | 0.3439(6) | 0.1601(2) | 0.3407(13) |
| **B (Å$^2$)** | 0.91(3) | 0.47(10) | 0.97(6) | 0.60(29) |
| **O22:** | - | *(4h) x½½* | - | *(4h) x½½* |
| **x** | - | 0.1642(7) | - | 0.1605(12) |
| **B (Å$^2$)** | - | 0.71(19) | - | 0.53(28) |
| **R$_{wp}$ (%)** | 7.86 | 7-84 | 8.31 | 8.06 |
| **R$_{Bragg}$ (%)** | 2.51 | 3.06 | 3.28 | 2.77 |

**Table I.** Structural parameters (lattice parameters, fractional coordinates and isotropic temperature factors) and reliability factors of the x-ray powder diffraction refinements at 80 and 295 K for La$_{0.45}$Sr$_{1.55}$MnO$_4$ and La$_{0.5}$Sr$_{1.5}$MnO$_4$.

| No. | Space group | Lattice vectors | Origin | Mn sites |
|---|---|---|---|---|
| 1 | Cmcm (No. 63) | (1,-1,0), (0,0,-1), (2,2,0) | (0,0,0) | 4a, 4c |
| 2 | Cmcm (No. 63) | (0,0,1),(-1,1,0),(-2,-2,0) | (0,0,0) | 4a, 4c |
| 3 | Cmca (No. 64) | (1,-1,0), (0,0,-1), (2,2,0) | (3/4,-1/4,1/4) | 8f |
| 4 | Cmca (No. 64) | (0,0,1),(-1,1,0),(-2,-2,0) | (-1/2,-1,1/2) | 8f |
| 5 | Cmmm (No. 65) | (0,0,1),(-1,1,0),(-2,-2,0) | (0,0,0) | 2a, 2d, 4l |
| 6 | Cccm (No. 66) | (0,0,1),(-1,1,0),(-2,-2,0) | (0,0,0) | 4b, 4c |
| 7 | Cmma (No. 67) | (0,0,1),(-1,1,0),(-2,-2,0) | (-1/2,-1,1/2) | 4g, 4g |
| 8 | Ccca (No. 68) | (0,0,1),(-1,1,0),(-2,-2,0) | (3/4,-1/4,1/4) | 8h |

**Table II.** Orthorhombic base centered subgroups (mmm point group) of the *I4/mmm* space group. The columns indicate the solution number, space group of the subgroup, lattice vectors and origin shift respect to the parent cell (high temperature phase) and the Wyckoff positions for the Mn sites in the subgroup.

| atom | site | x/a | y/b | z/c |
|---|---|---|---|---|
| Mn1 | 4a | 0 | 0 | 0 |
| Mn2 | 4c | 0 | 0.5000 | 3/4 |
| Sr/La1 | 8e | 0.6421 | 0 | 0 |
| Sr/La2 | 8g | 0.6421 | 0.5000 | 0.7500 |
| O11 | 8f | 0 | 0.7500 | 0.1250 |
| O12 | 8f | 0 | 0.7500 | 0.6250 |
| O21 | 8e | 0.8394 | 0 | 0 |
| O22 | 8g | 0.8394 | 0.5000 | 0.7500 |

**Table III.** Wyckoff positions and fractional coordinates for the undistorted structure of $La_{0.5}Sr_{1.5}MnO_4$ at 80 K in the *Cmcm* space group.

| No. | mode | atom | $\Delta x/a$ | $\Delta y/b$ | $\Delta z/c$ |
|---|---|---|---|---|---|
| 1 | $\Gamma_1^+(A1)$ | Sr/La1 | $-\delta_9$ | - | - |
|   |   | Sr/La2 | $-\delta_9$ | - | - |
| 2 | $\Gamma_1^+(A1)$ | O21 | $-\delta_{10}$ | - | - |
|   |   | O22 | $-\delta_{10}$ | - | - |
| 3 | $\Delta_2(Eu)$ | Mn1 | - | - | - |
|   |   | Mn2 | - | $-\delta_2$ | - |
| 4 | $\Delta_2(E)$ | Sr/La1 | - | - | - |
|   |   | Sr/La2 | - | $-\delta_7$ | - |
| 5 | $\Delta_2(B3u)$ | O11 | - | $+2\delta_5$ | $+\delta_5$ |
|   |   | O12 | - | $-2\delta_5$ | $-\delta_5$ |
| *6* | *$\Delta_2(B2u)$* | *O11* | - | *$-2\delta_1$* | *$+\delta_1$* |
|   |   | *O12* | - | *$+2\delta_1$* | *$-\delta_1$* |
| 7 | $\Delta_2(E)$ | O11 | - | - | - |
|   |   | O12 | - | $-\delta_8$ | - |
| 8 | $X_1^+(A1)$ | Sr/La1 | $-\delta_{11}$ | - | - |
|   |   | Sr/La2 | $+\delta_{11}$ | - | - |
| *9* | *$X_1^+(B2u)$* | *O11* | - | *$-2\delta_3$* | *$+\delta_3$* |
|   |   | *O12* | - | *$-2\delta_3$* | *$+\delta_3$* |
| 10 | $X_1^+(B3u)$ | O11 | - | $-2\delta_6$ | $-\delta_6$ |
|   |   | O12 | - | $-2\delta_6$ | $-\delta_6$ |
| *11* | *$X_1^+(A1)$* | *O21* | *$-2\delta_4$* | - | - |
|   |   | *O22* | *$+2\delta_4$* | - | - |

**Table IV.** Specific atomic motions for all the active modes. The three oxygen modes given rise to the low temperature orthorhombic phase are indicated in bold.

**FIGURE CAPTIONS.**

**Figure 1.** Rietveld refinement of La$_{0.45}$Sr$_{1.55}$MnO$_4$ at 80 K using the *Cmmm* space group.. Points and line correspond to the experimental and calculated pattern, respectively. The difference is plotted at the bottom below the bars marking the allowed reflections. The second row of bars corresponds to reflections arising from the LaSr$_2$Mn$_2$O$_7$ impurity (0.4% in wt). Inset: Detail of the x-ray patterns taken at 80 (squares) and 295 K (circles) showing a peak splitting. The superscript in the reflection index indicates the orthorhombic $c_t \times \sqrt{2}\, a_t \times \sqrt{2}\, a_t$ cell.

**Figure 2.** Temperature dependence of the lattice parameters for La$_{0.5}$Sr$_{1.5}$MnO$_4$ (circles) and La$_{0.45}$Sr$_{1.55}$MnO$_4$ (squares) samples. The subscripts *t* and *o* refers to tetragonal (*I4/mmm*) and orthorhombic (*Cmmm*) cells, respectively. The dotted line is a guide for the eyes marking the expected zone of the structural transition. The superscript in the lattice parameters refers to orthorhombic $c_t \times \sqrt{2}\, a_t \times \sqrt{2}\, a_t$ cell.

**Figure 3.** Atomic shifts for the equatorial oxygen atoms associated to the condensation of the modes indicated in each figure. For the sake of clarity, only Mn and equatorial oxygen atoms in a single layer are plotted as small (red for Mn1 and yellow for Mn2 in the e-version) and big balls (blue in the e-version), respectively. Oxygen displacements are indicated by arrows.

**Figure 4.** Intensities of $(3/2,3/2,0)_t$ (top panel) and $(5/2,5/2,0)_t$ (bottom panel) reflections as a function of the energy at the Mn K edge in the $\sigma$-$\sigma$' polarization channel at different values of the azimuth angle. The $\sigma$ polarization vector is parallel to $(-1,1,0)_t$ or $(1,-1,0)_t$ direction for $\varphi=0°$ and parallel to the $(0,0,1)_t$ one for $\varphi=-90°$.

**Figure 5**. Energy dependence of $(5/2, 5/2, 0)_t$ reflection in the $\sigma$-$\pi$' channel at different $\varphi$ values. For comparison, the intensity of $(3/2, 3/2, 0)_t$ in the same polarization channel at $\varphi=-45°$ is plotted in the same scale (grey solid line). The inset shows the intensity variation of the main resonance of $(5/2,5/2,0)_t$ as a function of $\varphi$ (open circles). The black solid line is an eye guide following an expression α $sin^2\varphi cos^2\varphi$.

**Figure 6**. Left y-scale: Energy dependence of $(h/4,h/4,0)_t$ forbidden reflections for $h=$ 5 (black solid line), 7 (circles) and 9 (continuous line) in the $\sigma$-$\pi$' channel at $\varphi=0°$. Right y-scale: Off-specular $(7/4,9/4,0)_t$ reflection in the $\sigma$-$\sigma$' channel (black line). Inset: azimuth evolution of the main resonance intensity in $(7/4\ 7/4\ 0)_t$ (triangles). The solid line is an eye guide following an expression $\alpha\ cos^2\varphi$.

**Figure 7.** Temperature dependence of the integrated intensity of $(3/2,3/2,0)_t$ superlattice (filled circles) and $(9/4,9/4,0)_t$ forbidden (open circles) reflections, both only present below $T_{CO} \sim 230$ K.

**Figure 8.** Energy dependence of superlattice and forbidden reflections calculated following the semi-empirical model (lines) compared to experimental RXS spectra (symbols): (a) $(3/2,3/2,0)_{t,\sigma-\sigma'}$ for $\varphi=0°$ (solid line, filled circles) and $\varphi=-90°$ (dotted line, open circles); (b) $(5/2,5/2,0)_{t,\sigma-\sigma'}$ for $\varphi=0°$ and $\varphi=-90°$, lines and symbols meaning as in (a); (c) $(5/2, 5/2, 0)_{t,\sigma-\pi'}$ for $\varphi=-45°$; and (d) $(7/4, 7/4, 0)_{t,\sigma-\pi'}$ at $\varphi=0°$.

**Figure 9.** Comparison between the theoretical RXS spectra (continuous line) and the experimental ones (dotted lines). Upper panel shows the polarized XANES spectra along the three main crystallographic directions. The other three panels shows the energy dependence of the $(3/2,3/2,0)_{t,\sigma-\sigma'}$, $(5/2,5/2,0)_{t,\sigma-\sigma'}$ and $(7/4,7/4,0)_{t,\sigma-\pi}$ reflections at the azimuth angles shown.

**Figure 10.** Oxygen coordination (red spheres) and interatomic Mn-O distances around Mn1 (violet sphere) and Mn2 (green sphere) in the *Cmcm* ordered phase. The picture does not follow the real scale and distortions of the octahedra have been exaggerated to facilitate their interpretation.

**Figure 1**

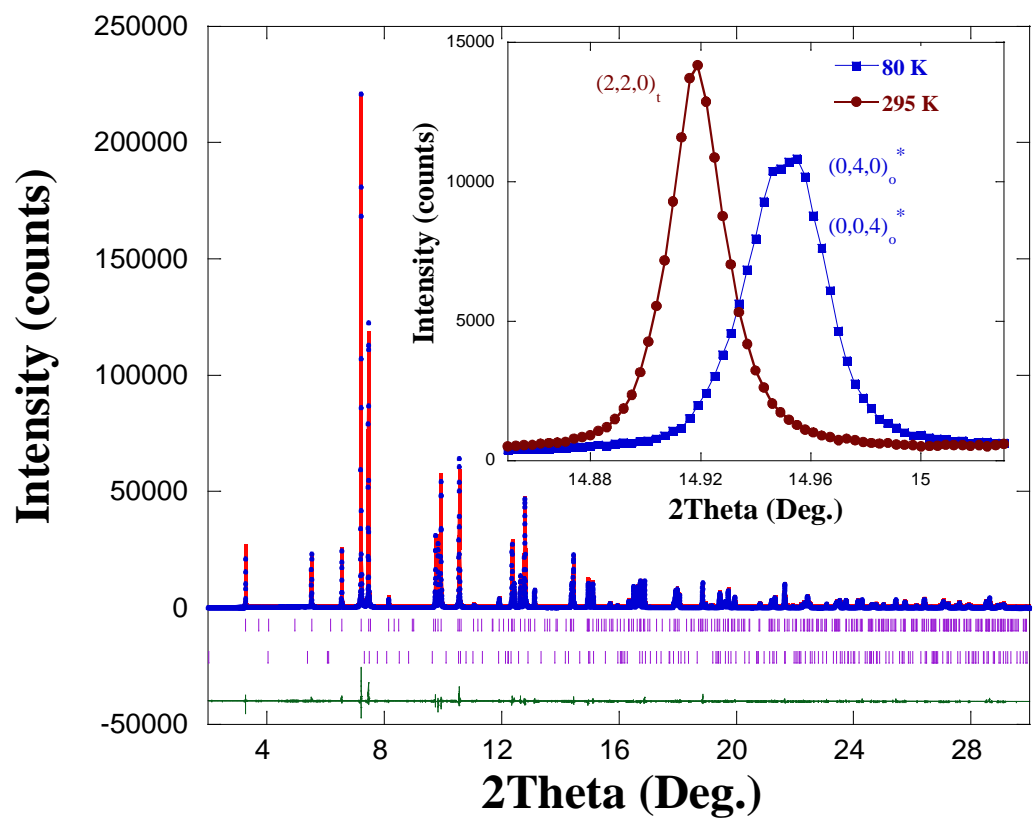

**Figure 2**

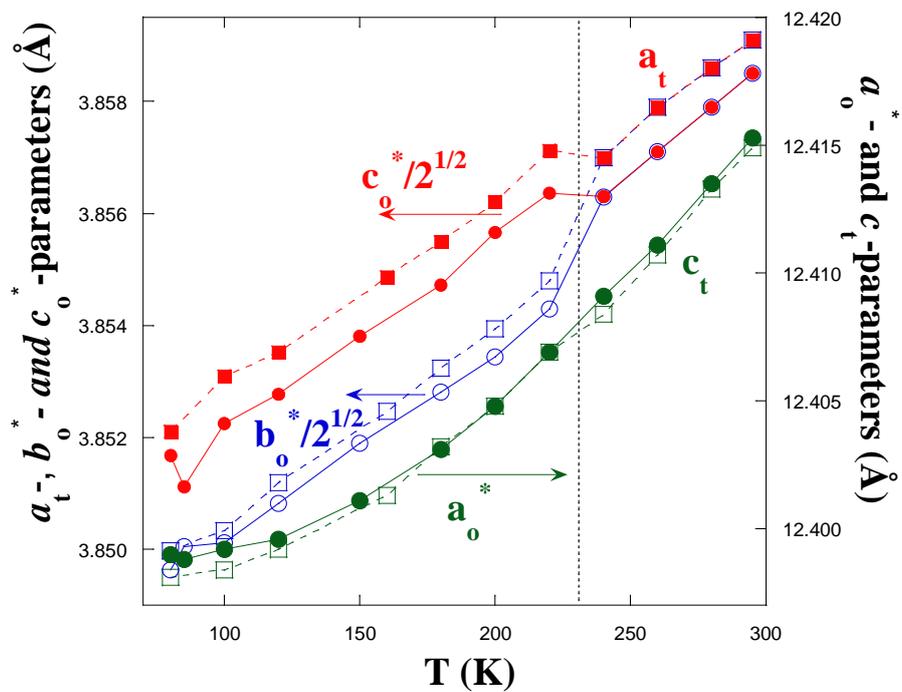

**Figure 3**

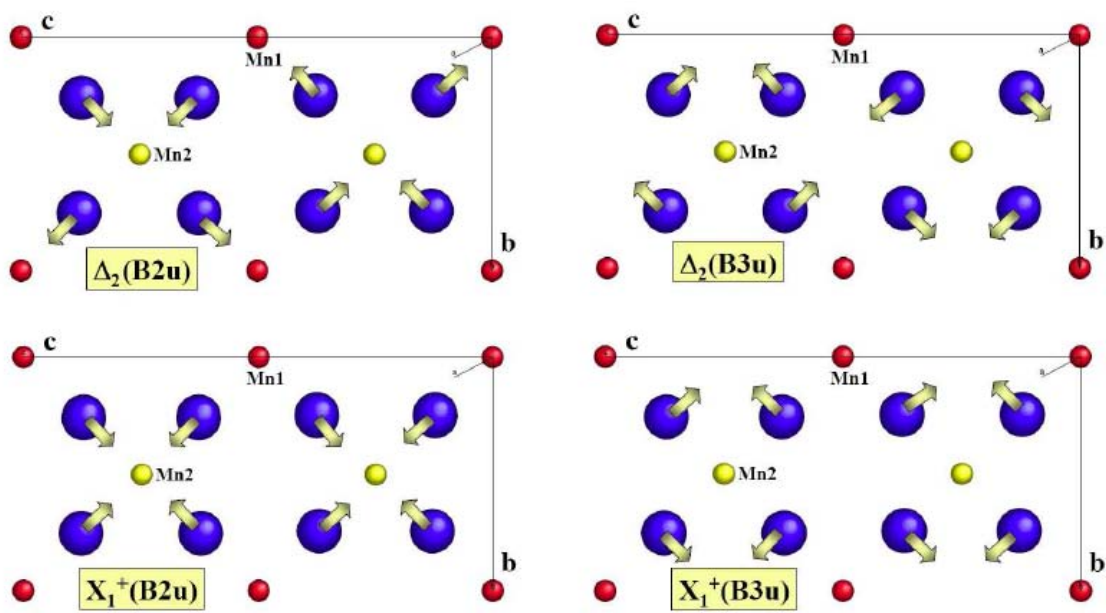

**Figure 4**

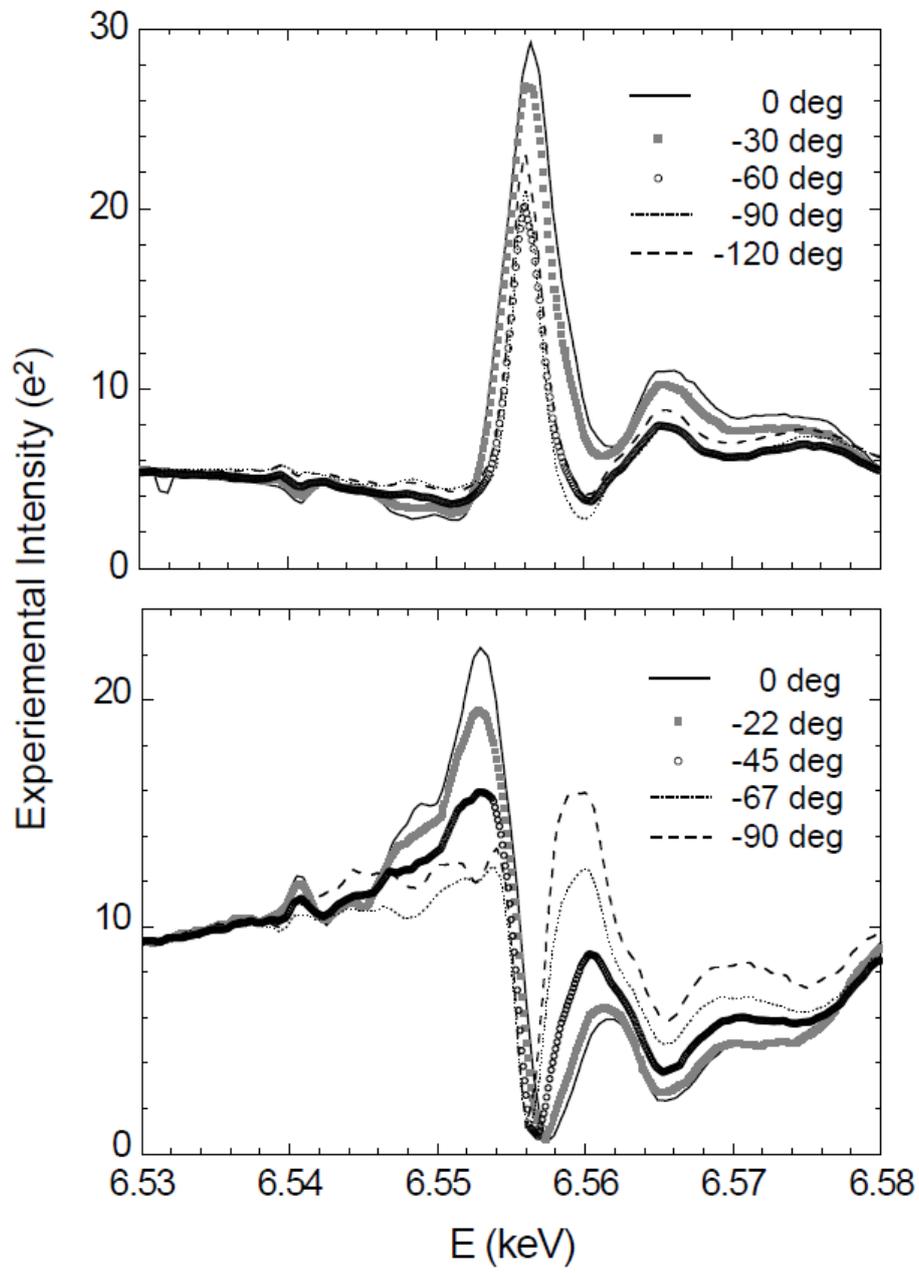

**Figure 5**

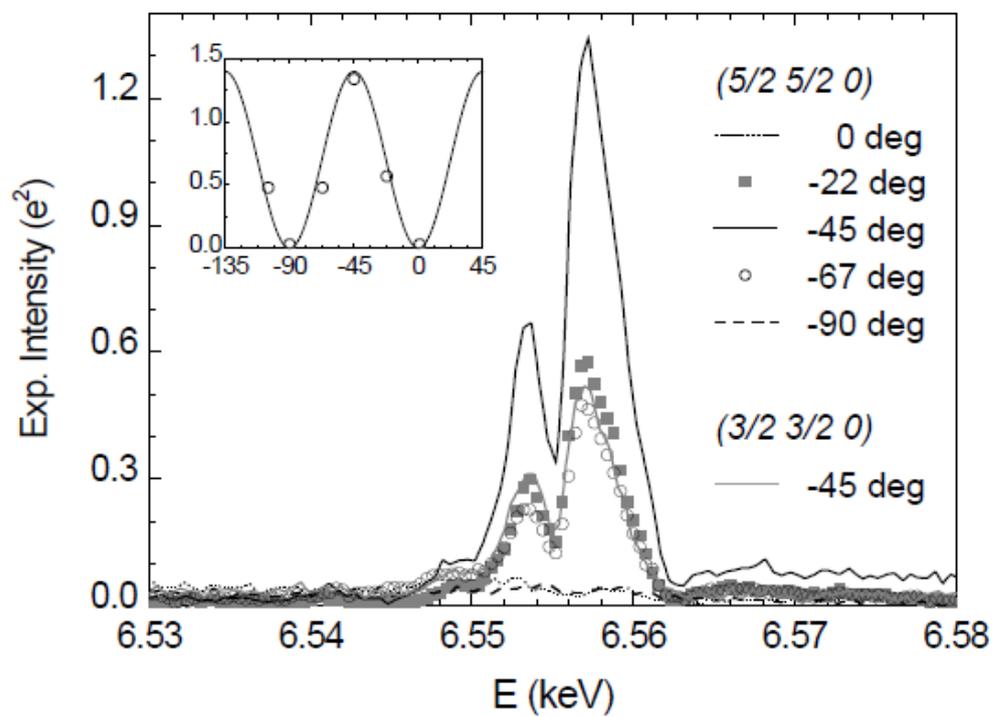

**Figure 6**

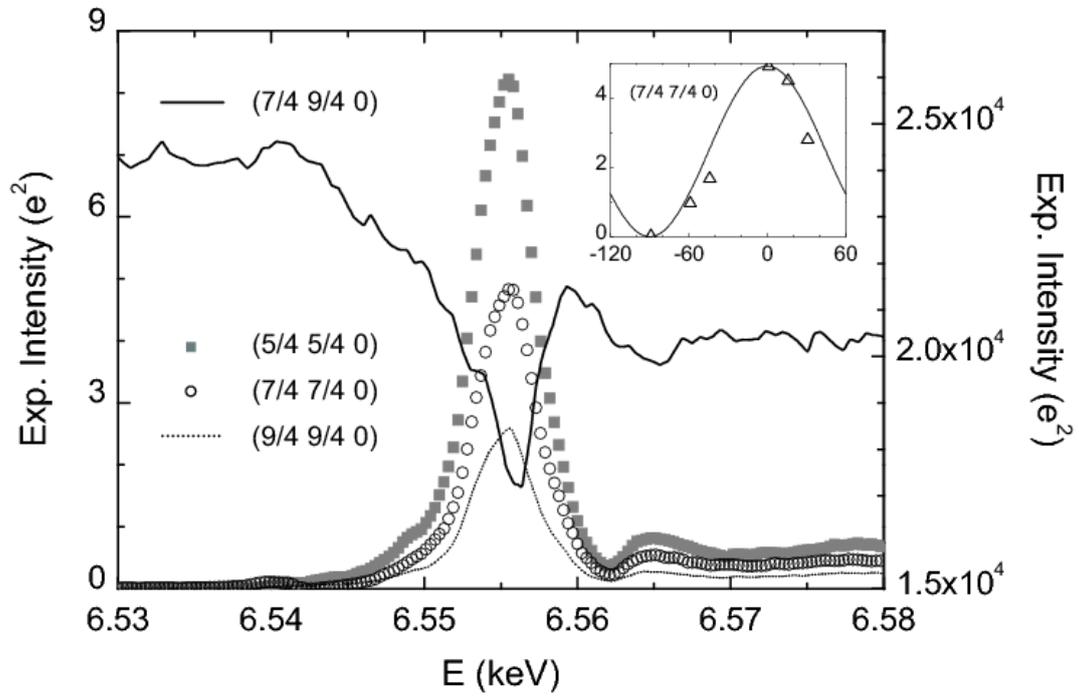

**Figure 7**

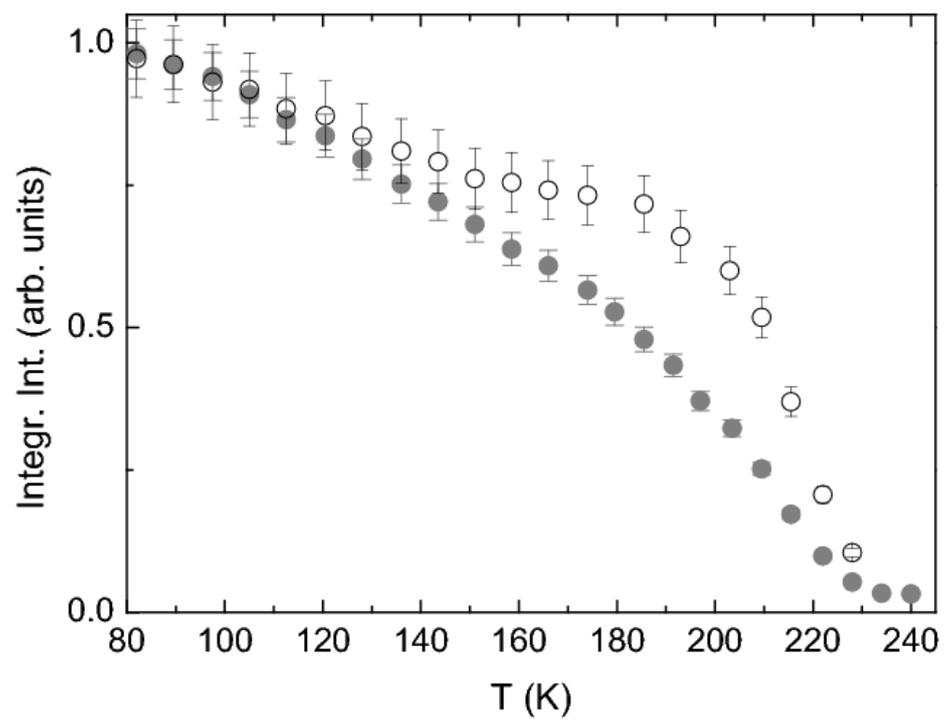

**Figure 8**

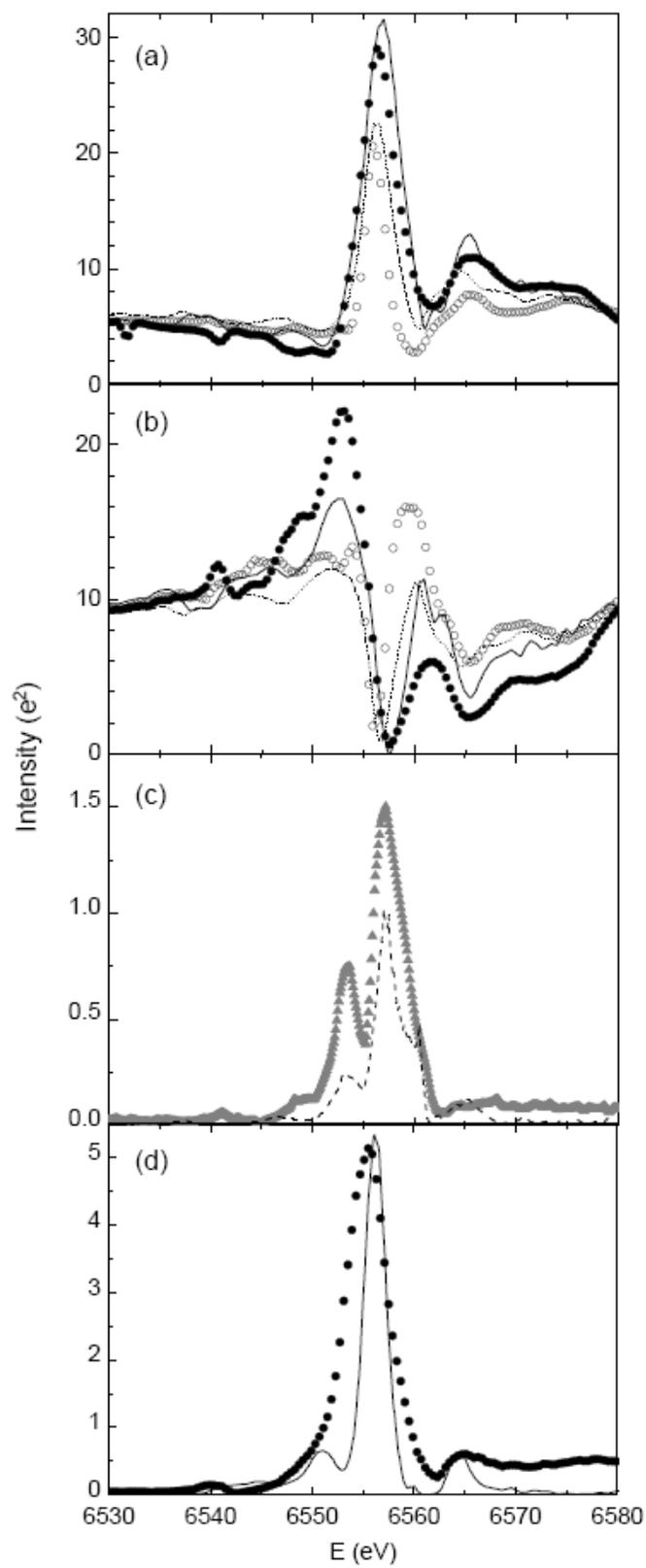

**Figure 9**

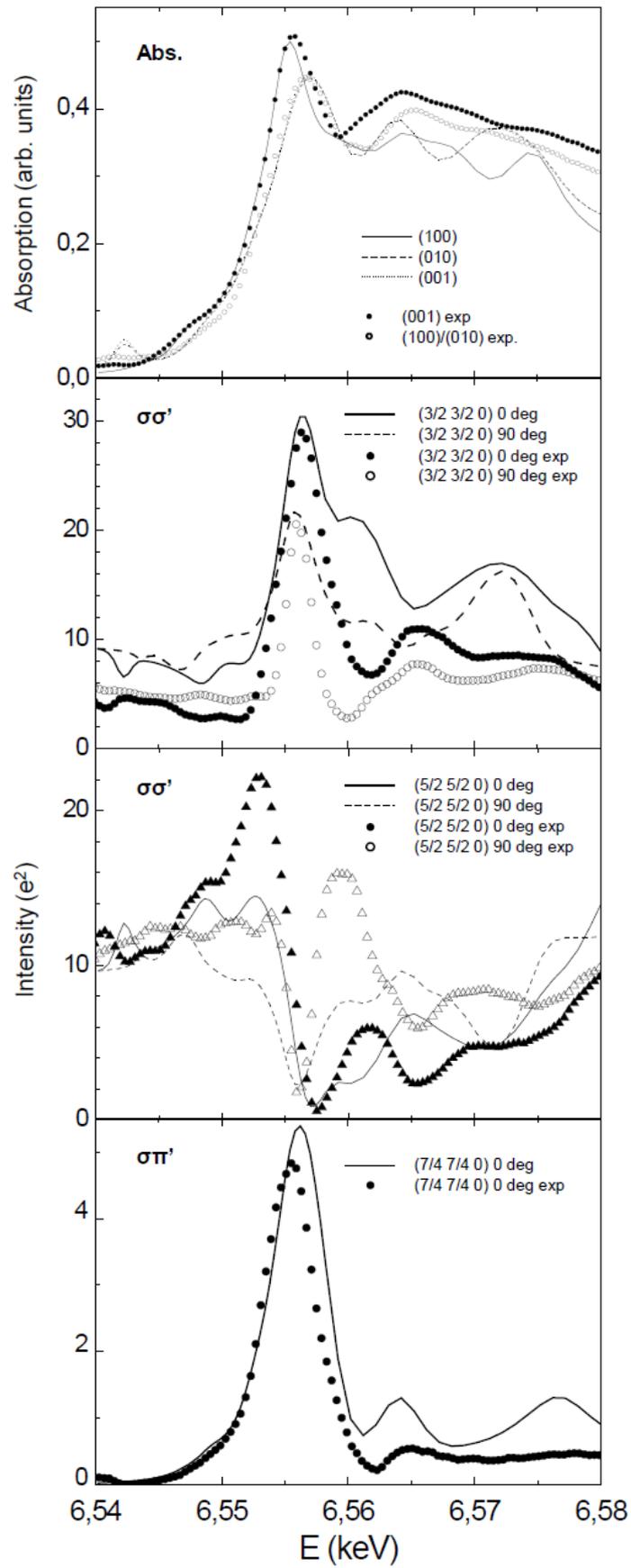

**Figure 10**

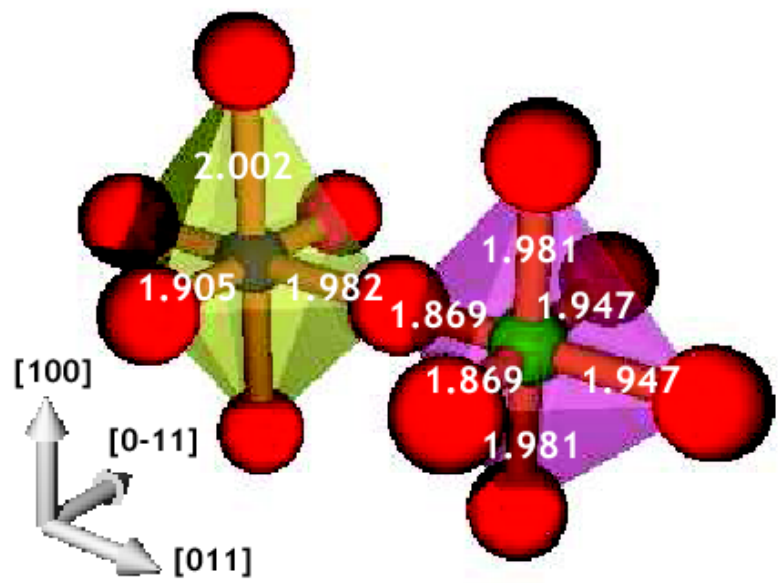